\documentclass[journal]{IEEEtran}

\usepackage{cite}
\usepackage{amsmath,amssymb,amsfonts}
\usepackage{algorithm,algorithmic}
\usepackage{algorithmic}
\usepackage{graphicx}
\usepackage{textcomp}
\usepackage{xcolor}
\usepackage{lineno}
\usepackage{mathtools}
\usepackage{csquotes}

\usepackage{subcaption}
\usepackage{lineno}

\usepackage{blindtext}

\ifCLASSINFOpdf
\else
\fi

\begin{document}
	
	\title{\huge  A Zero Trust Framework for Realization and Defense Against Generative AI Attacks in Power Grid}

	\author{
		\IEEEauthorblockN{\normalsize Md. Shirajum Munir\textsuperscript{1}, Sravanthi Proddatoori\textsuperscript{2}, Manjushree Muralidhara\textsuperscript{2}, Walid Saad\textsuperscript{3}, Zhu Han\textsuperscript{4}, and Sachin Shetty\textsuperscript{5}  }
		\IEEEauthorblockA{\normalsize \textsuperscript{1}School of Cybarsecurity, \textsuperscript{2}Dept. of CS, \textsuperscript{5}Dept. of ECE, Old Dominion University, Norfolk, VA 23529, USA\\
			\textsuperscript{3}Electrical and Computer Engineering, Virginia Tech, Arlington, VA 22203, USA\\
			\textsuperscript{4}Electrical and Computer Engineering, University of Houston, Houston, TX 77004, USA\\
			Email: mmunir@odu.edu; sprod002@odu.edu; mmura001@odu.edu; walids@vt.edu; hanzhu22@gmail.com; sshetty@odu.edu
		}
	\thanks{This work is supported in part by the DoD Center of	Excellence in AI and Machine Learning (CoE-AIML) under Contract Number W911NF-20-2-0277 with the U.S. Army Research Laboratory, National Science Foundation under Grant No. 2219742 and Grant No. 2131001, the Office of Naval Research (ONR) MURI Grant N00014-19-1-2621, VIRGINIA INNOVATION PARTNERSHIP CORPORATION Grant No 230849, the Commonwealth Cyber Initiative under contract number HC-3Q24-049, an investment in the advancement of cyber R\&D, innovation, and workforce development.}
	\thanks{©2024 IEEE. Personal use of this material is permitted. Permission from IEEE must be obtained for all other uses, in any current or future media, including reprinting/republishing this material for advertising or promotional purposes, creating new collective works, for resale or redistribution to servers or lists, or reuse of any copyrighted component of this work in other works.
}
	}

	\markboth{COPYRIGHT SUBJECT TO BE RESERVED TO IEEE}{}

	\maketitle
	\begin{abstract}
		Understanding the potential of generative AI (GenAI)-based attacks on the power grid is a fundamental challenge that must be addressed in order to protect the power grid by realizing and validating risk in new attack vectors. In this paper, a novel zero trust framework for a power grid supply chain (PGSC) is proposed. This framework facilitates early detection of potential GenAI-driven attack vectors (e.g., replay and protocol-type attacks), assessment of tail risk-based stability measures, and mitigation of such threats. First, a new zero trust system model of PGSC is designed and formulated as a zero-trust problem that seeks to guarantee for a stable PGSC by realizing and defending against GenAI-driven cyber attacks. 
		Second, in which a domain-specific generative adversarial networks (GAN)-based attack generation mechanism is developed to create a new vulnerability cyberspace for further understanding that threat. Third, tail-based risk realization metrics are developed and implemented for quantifying the extreme risk of a potential attack while leveraging a trust measurement approach for continuous validation. Fourth, an ensemble learning-based bootstrap aggregation scheme is devised to detect the attacks that are generating synthetic identities with convincing user and distributed energy resources device profiles. Experimental results show the efficacy of the proposed zero trust framework that achieves an accuracy of $95.7\%$ on attack vector generation, a risk measure of $9.61\%$ for a $95\%$ stable PGSC, and a $99\%$ confidence in defense against GenAI-driven attack. 
   
	\end{abstract}

	\section{Introduction}
	Power grid supply chain (PGSC) cybersecurity is necessary to the infrastructure that provides electrical power to homes, businesses, and critical facilities. The PGSC infrastructure is expected to deploy around $30$-$40$ billion distributed energy resource (DER) devices such as renewable energy sources, consumers, prosumers, generators, electric vehicles (EV), EV charging stations, and so on by $2025$ to meet an envisioned $40\%$ energy cost reduction by $2050$ \cite{CySe_1, CySe_4, CySe_6, no_of_devices}. The rigorous expansion of diversified DERs brings indispensable cyber challenges for power grid operations \cite{CySe_1, CySe_4, CySe_6, Munir_WSC_2023} by creating a large surface. Additionally, artificial intelligence (AI) can induce adversarial attacks on PGSC \cite{dash2023_GAN_4, ben2023federated, dash2023_GAN_5}.  
	
	Generative artificial intelligence (GenAI) models such as generative adversarial networks (GAN) \cite{goodfellow_GAN_1, chaccour2022less, TWC_SAAD_GAN_7, IoT_SAAD_GAN_8} offer significant benefits in data augmentation and reconstruction. Therefore, GANs can expand the of cyber attack vectors in the power grid by generating synthetic identities with convincing user and DER device profiles \cite{dash2023_GAN_4, dash2023_GAN_5}. In particular, GenAI can create new attack vectors for launching replay attacks by generating observed control message parameters such as the reaction time of participants, nominal power consumed, price elasticity coefficient \cite{UCI_Grid_Stability_Data} and their pattern from the trusted DERs. GenAI can also imitate the broadcast data distribution of DERs such as communication data packet, packet size, IP, port, demand-response energy data, and so on for introducing protocol-type attacks in PGSC \cite{WUSTL_data_2}. These types of attack vectors have not been included in DER security standard IEEE 1547 \cite{photovoltaics2018ieee1547}.
	Clearly, advances in GenAI can lead to \emph{novel attack surfaces} that, in turn, introduce new vulnerabilities and risks to the power grid, which can potentially lead to \emph{1) unauthorized parties gaining access due to de-synchronized control and communication messages by protocol attack, and 2) power outage, energy theft, and money fraud are caused by replay attacks on nominal power consumed and price elasticity coefficient of DERs.} 
	
	In order to defend against these new vulnerabilities, it is essential to address several unique challenges that include: 
	\begin{itemize}
		\item  Generation of potential attacks that can be created by GenAI in order to understand the potential vulnerabilities in advance.
		\item Design of tail and risk-based reliability measure and trust metrics to analyze the worst-case vulnerabilities of various energy DERs control and communication messages for low latency recovery, and adaptation of energy grid behavior changes. 
		\item Moving from classical \emph{trust and verify approaches} into a zero-trust regime built on the paradigm of \emph{never trust and always verify} which effectively identify, explain, and defend any disrupted events carried on by GenAI in PGSC.
	\end{itemize}
	
	The main contribution of this paper is to address the above technical challenges by proposing a \emph{zero trust framework} for risk measuring and defense against GenAI-driven attacks on the PGSC. Towards developing this framework, we make the following key contributions:
	\begin{itemize}
		\item We design a new zero trust system model of PGSC and formulate a joint optimization problem for generating novel attack surfaces, measuring risk, and defense against the generated control/status message of DERs. 
		
		\item We develop a domain-specific GAN mechanism for potential vulnerability creation. Here, the main novelty is the capability of generating new attack vectors for further understanding by modeling generative adversarial networks for generating synthetic identities that convincingly mimic the device profiles of legitimate users and DER device profiles. 
		
		\item We develop tail-based reliability metrics for realizing the risk of potential attack. Then, we propose a trust quantification approach for continuous validation on understanding the underlying risk of DERs'. 
		
		\item We devise a defense strategy for GenAI-driven attacks on PGSC by leveraging an ensemble learning method (i.e., a bootstrap aggregation (bagging) mechanism) for solving a random forests (RF) regression problem.
		
		\item The performance of the developed zero trust framework is validated by leveraging two state-of-the-art PGSC datasets. Our experimental analysis shows that the proposed zero trust framework can successfully generate control/status (about $95.7\%$),  quantify extreme risk (around $9.61\%$) for PGSC stability parameters with a $95\%$ confidence (trust), and achieve around $99\%$ accuracy for GenAI-driven attacks detection on PGSC.    
	\end{itemize}

	\begin{table}[t!]
		\caption{Summary of notations.}
		\begin{center}
				\begin{tabular}{|p{1.5cm}|p{6cm}|}
					\hline
					\textbf{Notation}&{\textbf{Description}} \\
					\hline
					$\mathcal{I}$ & A set of DRE \\
					\hline 
					$q_i(t)$& Power (i.e., +ve for generator, -ve for consumption)\\
					\hline
					$\boldsymbol{x}_{it}$ & Control/status message\\
					\hline 
					$\Theta_{i}$ & Rotor angle \\
					\hline
					$\beta_i$ & Damping constraint \\
					\hline
					$\alpha_{ij}$ & Coupling strength between $i$ and $j$  \\
					\hline
					$\hat{q_i}$  & Power  \\
					\hline
					$\Phi_i$  & PGSC market elasticity  \\
					\hline
					$\tau_i$  & Response delay \\
					\hline
					$G_\theta$ & Generator \\
					\hline
					$D_\phi$ & Discriminator \\
					\hline 
					$\eta \in (0,1)$ & CVaR significant probability \\
					\hline
					$\xi$ & CVaR confidence level\\
					\hline
				\end{tabular}
				\label{tab1_SummaryofNotation}
			\end{center}
		\end{table}
		
		\section{System Model for Realizing GenAI-Driven Attacks in PGSC}
		\begin{figure}[!t]
			\centerline{\includegraphics[scale=.45]{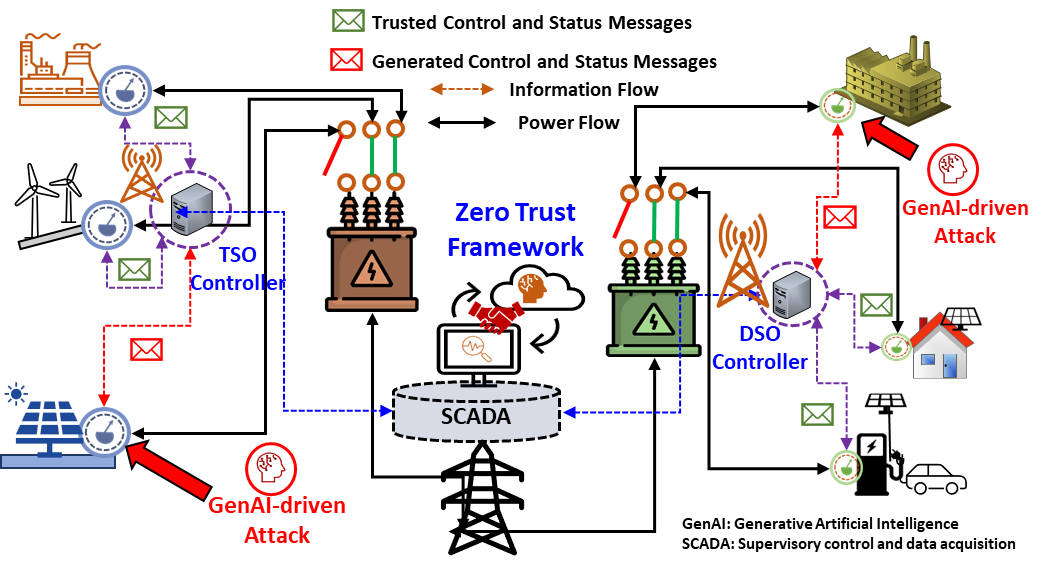}}
			\caption{A system model of a zero trust framework for risk realization and defense against GenAI attacks on the PGSC.}
			\label{System_model}
			\vspace{-2mm}
		\end{figure}
		
		We consider a power grid supply chain equipped with a set  $\mathcal{I}$ of $I$ DERs such as generators, consumers, and prosumers (as seen in Figure \ref{System_model}). 
		In our system, we consider finite, continuous time, such that each time slot $t \in (0, T)$. 
		Therefore, at time slot $t$, each DER $i \in \mathcal{I}$ can generate $q_i(t)$ (i.e., $q_i(t)$ is a positive value) or consume $q_i(t)$ (i.e., $q_i(t)$ is negative value) power. In this PGSC, supervisory control and data acquisition (SCADA) systems monitor and orchestrate the power grid operation while the transmission system operator (TSO) and distribution system operator (DSO) assist in transferring and distributing energy, respectively. In particular, TSO brings energy from production to the main grid while DSO distributes it to the end users such as consumers.
		
		At the time $t$, each DER $i \in \mathcal{I}$ can exchange SCADA control and status message $\boldsymbol{x}_{it}$ with the SCADA system. Consequently, DER $i \in \mathcal{I}$ can send and receive control message vector $\boldsymbol{x}_{it} = (a_{it}, b_{it}, c_{it}, d_{it}, e_{it})$ to execute operational command (e.g., energy supply, grid health maintenance, connect/disconnect from the main grid, etc). Each message $\boldsymbol{x}_{it}$ contains send packet $a_{it}$, send packet size $b_{it}$, number of packets source to destination $c_{it}$, number of packet destination to source $d_{it}$, and total received packets $e_{it}$. Fake or generated control messages create a major risk for cyber vulnerabilities by executing protocol and replay attacks in PGSC. In particular, the reconstruction capability of GenAI introduces a high risk of protocol and replay attacks in PGSC. Thus, PGSC is potentially under the high risk of cyber vulnerabilities that may create power outages, grid health, information theft, unstable market, and so on. 
		We will hence introduce a novel system model for identifying the cyber vulnerabilities risk of the potential GenAI-driven cyber attacks for assuring a stable PGSC. 
		\vspace{-2mm}
		\subsection{Power Grid Supply Chain Stability Model}
		In our model, each DER $i$ can transfer energy to other DERs in set $\mathcal{I}$. All DERs in $\mathcal{I}$ are equipped with oscillators. Now, for transferring energy from DER $i$ to DER $j \in \mathcal{I}, i \ne j$, we define a coupling strength $\alpha_{ij}$, a rotor angle $\Theta_{i}$, and a damping constraint $\beta_i$ for DER $i$. We can now define the dynamics of power transmission by an oscillator model \cite{IEEEhowto:model_base_eq} in PGSC,  
		\vspace{-2mm}
		\begin{equation} \label{eq:Oscillator_model}
			\frac{d^2\Theta_i}{dt^2} = q_i -\beta_i \frac{d\Theta_i}{dt} + \sum_{j=1, j \ne i}^{I} \alpha_{ij} \sin (\Theta_j - \Theta_i),
			\vspace{-2mm}
		\end{equation}
		where $q_i$ represents the power. Therefore, the power transfer between DER $i$ to DER $\forall j \in \mathcal{I}, i \ne j$ is relay on the time derivative. Therefore, for produced/supply power $\hat{q_i}$, the oscillator model \eqref{eq:Oscillator_model} can be presented as follows \cite{IEEEhowto:Decentral_base}:   
		\vspace{-2mm}
		\begin{equation} \label{eq:fn_of_freq_deviation}
			\hat{q_i} (t) = q_i -\Phi_i \frac{d\Theta_i}{dt} (t),
			\vspace{-1mm}
		\end{equation}
		where $\Phi_i$ is the elasticity of DER $i$ and $\Phi_i$ is proportional to energy market elasticity \cite{IEEEhowto:Decentral_base}. Further, the rotation reference of angular frequency deviation $\frac{d\Theta_i}{dt}$ depends on the power grid architecture such as $2 \pi \times 50$ Hz or $2 \pi\times 60$ Hz. Consequently, the potential supply chain instability is induced by a response delay $\tau_i$ of each DER $i \in \mathcal{I}$ (i.e., generator and consumer in PGSC). Then, we can present transmission power $\hat{q_i} (t)$ as $\hat{q_i} (t-\tau)$, where $\tau$ is the response delay. We can now derive a new  oscillator model using \eqref{eq:Oscillator_model} and \eqref{eq:fn_of_freq_deviation}:    
		\vspace{-1mm}
		\begin{equation} \label{eq:Oscillator_model_with_tao}
			\frac{d^2\Theta_i}{dt^2} = q_i -\beta_i \frac{d\Theta_i}{dt} + \sum_{j=1}^{I} \alpha_{ij} \sin (\Theta_j - \Theta_i) -\Theta_i \frac{d\Theta_i}{dt} (t-\tau).
		\end{equation}         
		Clearly, the stability of PGSC $s_i(t) \approx \frac{d^2\Theta_i}{dt^2}$ relies on the physical behavior and control message $\boldsymbol{x}_{it}$ of each DER $i \in \mathcal{I}$. For measuring the PGSC stability in a finite time interval length of $T$, we can write grid stability as follows: 
		\vspace{-2mm}
		\begin{equation} \label{eq:Oscillator_model_with_freq_delay}
			\begin{split}
				s_i(t) \approx \frac{d^2\Theta_i}{dt^2} = q_i -\beta_i \frac{d\Theta_i}{dt} + \sum_{j=1}^{I} \alpha_{ij} \sin (\Theta_j - \Theta_i) - \\\frac{\Phi_i}{T} \int_{{t-T}}^{t} \frac{d\Theta_i}{dt} (t'-\tau) dt'.
			\end{split}
		\end{equation}
		$s_i(t)$ can be used to assess whether the PGSC is stable or not, For instance, a positive value of $s_i(t)$ means the PGSC is linearly unstable. Therefore, GenAI can manipulate and create fake parameters (e.g., rotor angle $\Theta_{i}$, damping constraint $\beta_i$, elasticity $\Phi_i$, etc) of a control message $\boldsymbol{x}_{it}$. To assess the risk of GenAI-driven attacks, in our model, we use GAN to analyze the capability of a new attack surface in PGSC.  
		\vspace{-3mm}
		\subsection{GAN for Identifying GenAI-driven Attack Vectors on PGSC}
		\label{sub_Gan}
		We use GAN \cite{goodfellow_GAN_1} to uncover the new attack surface on PGSC. We specifically leverage GAN to reproduce the PGSC control and status messages $\boldsymbol{x}_{it}$ to examine the risk of cyber vulnerabilities and power grid instability. Considering a likelihood-free generator $G_\theta$ can generate operational control message $\boldsymbol{x}_{it}$, where $\theta$ denotes learning parameters. We introduce a discriminator $D_\phi$ with parameters $\phi$. 
		Therefore, generator $G_\theta$ can generate control message $\boldsymbol{x}_{it}$ from sample $\boldsymbol{z}_{it}$ based on some latent variables, where intuitively, $\boldsymbol{z}_{it}$ is a noise vector. We define $y_{it}$ as a decision variable that discriminator $D_\phi$ uses to predict whether $\boldsymbol{x}_{it}$ is a generated control message or not. 
		Consequently, a control message generator $G_\theta$ minimizes the residual between two sample distribution $P_{\mathbb{X}} \approx P_{\theta}$ while discriminator $D_\phi$ maximizes the distance distribution of $P_{\mathbb{X}}$ and $P_{\theta}$, where $\mathbb{X}$ is a given distribution of the DER control message. We can write the GAN model as follows \cite{goodfellow_GAN_1}:
		\begin{equation} \label{eq:GAN_1}
			\begin{split}
				\underset{\theta}\min \; \underset{\phi} \max \; U (G_\theta, D_\phi) = \underset{\theta}\min \; \underset{\phi} \max  \; \mathbb{E}_{\boldsymbol{x}_{it}\sim P_{\mathbb{X}}} \\ \big[ \log D_\phi(\boldsymbol{x}_{it}) \big] +  \mathbb{E}_{\boldsymbol{z}_{it}\sim P_{\boldsymbol{z}_{it}}}\big[ \log (1 -  D_\phi(G_\theta (\boldsymbol{z}_{it})) )\big].
			\end{split}
		\end{equation}
		
		In \eqref{eq:GAN_1}, for a given generator $G_\theta$, the discriminator $D_\phi$ is maximizing the objective with respect to parameters $\phi$. The discriminator $D_\phi$ then performs the role of a binary classification decision $y_{it}$ (i.e., whether the control message is original or fake) on $\boldsymbol{x}_{it}\sim P_{\mathbb{X}}$. We define $P_{\mathbb{X}} (\boldsymbol{x}_{it})$ and $P_G (\boldsymbol{x}_{it})$ as, respectively, the probability of an actual and generated control message. Hence, the discriminator $D_\phi$ can be written as follows:  
		\vspace{-2mm}
		\begin{equation} \label{eq:GAN_dis_detailed}
			\begin{split}
				\hat{D}_\phi(\boldsymbol{x}_{it}| G_{\theta}) = \frac{P_{\mathbb{X}} (\boldsymbol{x}_{it})}{P_{\mathbb{X}} (\boldsymbol{x}_{it}) + P_{G} (\boldsymbol{x}_{it})}.
			\end{split}
		\end{equation}
		We can observe the probability of generated control message of DER $i$ at time $t$ by estimating \eqref{eq:GAN_dis_detailed}. Therefore, a generated control message $\boldsymbol{x}_{it}$ has significantly increased the risk of cyber vulnerability and energy market instability in the PGSC. The generated control message can execute a replay and protocol attack in PGSC. In particular, the GAN can reproduce a copy of a DER control message such as send packet $a_{it}$, send packet size $b_{it}$, number of packets source to destination $c_{it}$, number of packet destination to source $d_{it}$, and total received packets $e_{it}$ while capable of manipulating rotor angle $\Theta_{i}$, damping constraint $\beta_i$, elasticity $\Phi_i$, and so on. 
		
		In this work, we develop a \emph{Zero trust framework for risk realization and defense against GenAI-driven cyber attacks} in the PGSC. Therefore, we consider extreme value theory such as conditional-value-at-risk (CVaR) \cite{Saad_THz_reliability, Munir_CvaR2, Munir_CvaR1} to realize AI-driven cyber vulnerabilities in PGSC. 

		\section{GenAI-driven Vulnerability Risk Assessment Problem Formulation of PGSC}
		Next, we formulate a zero trust risk assessment problem to understand GenAI-driven cyber vulnerability on PGSC. We quantify the tail risk of cyber attacks by leveraging the concept of CVaR \cite{Saad_THz_reliability, Munir_CvaR1, rockafellar2002conditional}. In particular, we formulate a residual minimization problem for quantifying tail risk of a AI-generated control message $\boldsymbol{x}_{it}$ at DER $i \in \mathcal{I}$ while satisfying CVaR confidence level $\xi$. We consider $h(\boldsymbol{x}_{it}, \xi)$ is a probability distribution of trustworthy control message while $\xi$ can be a cut-off point of a risk deviation function $\Upsilon(\boldsymbol{x}_{it},\boldsymbol{z})$, where $\boldsymbol{z}$ represents latent variables of GAN (see detailed in section \ref{sub_Gan}). Thus, for a CVaR confidence $\xi$, a cumulative distribution function (CDF) can be calculated as follows \cite{rockafellar2002conditional}: 
		\begin{equation} \label{eq:cvar_threshold_cdf}
			h(\boldsymbol{x}_{it}, \xi) = \int_{\Upsilon(\boldsymbol{x}_{it},\boldsymbol{z}) \le \xi} P(\boldsymbol{z})d\boldsymbol{z},
		\end{equation}
		where $\xi$ is inversely proportional to $\Upsilon(\boldsymbol{x}_{it},\boldsymbol{z})$. In \eqref{eq:cvar_threshold_cdf}, $h(\boldsymbol{x}_{it}, \xi)$ becomes a nondecreasing and continuous function \cite{rockafellar2002conditional, Munir_CvaR1} because $\xi$ satisfies $\Upsilon(\boldsymbol{x}_{it},\boldsymbol{z}) \le \xi$. For a CVaR significant probability $\eta \in (0,1)$, we can define a random variable $\Psi_{\eta} (\boldsymbol{x}_{it})$ of control message $\boldsymbol{x}_{it}$. Therefore, we can define a value-at-risk quantification function $\xi_{\eta} (\boldsymbol{x}_{it})$ of control message $\boldsymbol{x}_{it}$ as follows:       
		\begin{equation} \label{eq:var_eq}
			\xi_{\eta} (\boldsymbol{x}_{it}) = \underset{\xi \in \mathbb{R}} \min \; h(\boldsymbol{x}_{it}, \xi) \ge \eta.
		\end{equation}
		
		We can estimate $\xi$ in \eqref{eq:var_eq} by satisfying $h(\boldsymbol{x}_{it}, \xi) \ge \eta$ and $\xi_{\eta} (\boldsymbol{x}_{it})$ becomes an upper-bound of tail risk on control message $\boldsymbol{x}_{it}$. Therefore, we can capture a conditional expectation of CVaR $\Psi_{\eta}(\boldsymbol{x}_{it})$ of AI generated control message $\boldsymbol{x}_{it}$ as follows:
		\begin{equation} \label{eq:cvar_eq}
			\begin{split}
				\underset{\xi \in \mathbb{R}} \min \; \frac{1}{(1-\eta)}\int_{P(\Upsilon(\boldsymbol{x}_{it},\boldsymbol{z})) \ge \xi_{\eta} (\boldsymbol{x}_{it})} \Upsilon(\boldsymbol{x}_{it},\boldsymbol{z}) P(\boldsymbol{z})d\boldsymbol{z},
			\end{split}
			\vspace{-4mm}
		\end{equation}
		where $P(\Upsilon(\boldsymbol{x}_{it},\boldsymbol{z})) \ge \xi_{\eta} (\boldsymbol{x}_{it}) = (1-\eta)$. Therefore, we can define the tail-risk realization objective $\Lambda_{\eta}(\boldsymbol{x}_{it}, \xi)$ as follows:
		\begin{equation} \label{eq:var_cvar_eq}
			\begin{split}
				\underset{\xi \in \mathbb{R}} \min \; \xi + \frac{1}{(1-\eta)} \int_{h(\boldsymbol{x}_{it}, \xi) \ge \xi} [h(\boldsymbol{x}_{it}, \xi) - \xi]^{+} P(\boldsymbol{z})d\boldsymbol{z}.
			\end{split}
		\end{equation}
		In CVaR formulation \eqref{eq:var_cvar_eq}, $[h(\boldsymbol{x}_{it}, \xi) - \xi]^{+}$ is positive and continuous since $h(\boldsymbol{x}_{it}, \xi)$ is a continuous function in \eqref{eq:cvar_threshold_cdf}. An approximate function of CVaR in \eqref{eq:var_cvar_eq} will be::
		\vspace{-4mm}
		\begin{equation} \label{eq:var_cvar_eq_aux}
			\begin{split}
				\hat{\Lambda}_{\eta}(\boldsymbol{x}_{it}, \xi) = 
				\underset{\xi, \boldsymbol{x}_{it}, y_{it}} \min \; \xi + \frac{1}{(1-\eta)} \frac{1}{|\mathcal{I}| T }\sum_{t=1}^{T} \sum_{i=1}^{|\mathcal{I}|} \Delta_{it},
			\end{split}
			\vspace{-4mm}
		\end{equation}
			where  $\Delta_{it} \ge (h(\boldsymbol{x}_{it}, \xi) - \xi)$ and $\Delta_{it} \ge 0$.
			Therefore, we formulate the risk-realization problem of GenAI-driven control message in PGSC as follows: 
			\begin{subequations}\label{Opt_1_1}
				\begin{align}
					\underset{\xi, \boldsymbol{x}_{it}, y_{it}} \min \;
					&\;  \xi + \frac{1}{(1-\eta)} \frac{1}{|\mathcal{I}| T }\sum_{t=1}^{T} \sum_{i=1}^{|\mathcal{I}|} \Delta_{it}  \tag{\ref{Opt_1_1}}, \\
					\text{s.t.} \quad & \label{Opt_1_1:const1} \Delta_{it} \ge (h(\boldsymbol{x}_{it}, \xi) - \xi), \Delta_{it} \ge 0, \\
					& \label{Opt_1_1:const2} \hat{D}_\phi(\boldsymbol{x}_{it}| G_{\theta}) \ge \frac{P_{\mathbb{X}} (\boldsymbol{x}_{it})}{P_{\mathbb{X}} (\boldsymbol{x}_{it}) + P_{G} (\boldsymbol{x}_{it})},\\
					& \label{Opt_1_1:const3} h(\boldsymbol{x}_{it}, \xi) \ge \eta, \eta \in (0,1),\\
					&\label{Opt_1_1:const4} s_i(t) \le 0, s_i(t) \in \boldsymbol{x}_{it}, s_i(t) \in (-1, 1), \\
					&\label{Opt_1_1:const5}  y_{it} \ge \omega_0 + \omega_1 z_{1i} +  \dots + \omega_N z_{Ni}, \forall z_{Ni} \in \boldsymbol{z}_{it},  \\
					& \label{Opt_1_1:const6} y_{it} \in \left\lbrace0,1 \right\rbrace, y_{it} \in \boldsymbol{y}, \forall i \in \mathcal{I}.
				\end{align}
		\end{subequations}
		The objective of \eqref{Opt_1_1} is to minimize the expected shortfall (i.e., mean-variance) with a given significant label of risk $\eta$ on a generated AI-driven control message in PGSC. Therefore, in \eqref{Opt_1_1}, we have three decision variables, CVaR cut-off point in long-tail distribution $\xi$, generated control message $\boldsymbol{x}_{it}$ of DER $i \in \mathcal{I}$, and binary decision variable $y_{it} \in \boldsymbol{y}$ to determine whether the control messages become fake or real. Constraint \eqref{Opt_1_1:const1} provides to an upper-bounded equivalent function of original objective \eqref{eq:var_cvar_eq_aux}. Constraint \eqref{Opt_1_1:const3} assigns a probability for determining an actual and generated control message of DER $i\in \mathcal{I}$ during the GAN fake message generation. Then, constraint \eqref{Opt_1_1:const3} ensures a certain significant level $\eta \in (0,1)$ (e.g., $0.95$) of tail risk for a generated AI-driven control message $\boldsymbol{x}_{it}$. Constraint  \eqref{Opt_1_1:const4} establishes a connection among the grid stability parameters such as rotor angle $\Theta_{i}$, damping constraint $\beta_i$, elasticity $\Phi_i$ of oscillator model \eqref{eq:Oscillator_model_with_freq_delay} to transfer energy. Constraint \eqref{Opt_1_1:const4} assures a stable PGSC by restricting $s_i(t)$ to negative values.
		Constraint \eqref{Opt_1_1:const5} establishes a relationship between the GAN's latent variables $\boldsymbol{z}$ and a regression weight $\omega$ for distinguishing $y_{it} \in \boldsymbol{y}$ among generated and original control message. Finally, constraint \eqref{Opt_1_1:const6} assures $y_{it}$ as a binary variable for each control message $\boldsymbol{x}_{it}$.
		
		The formulated zero trust problem \eqref{Opt_1_1} is to a combinatorial optimization problem due to the relationship among the corresponding constraints. Further, decision variables of the formulated problem \eqref{Opt_1_1} belong to both time and space domains while they are correlated. As a result, the formulated zero trust problem \eqref{Opt_1_1} is hard to solve in polynomial time complexity. Therefore, we propose a zero trust framework for extreme risk realization and defense against generated-AI driven attacks on PGSC. In particular, the proposed zero trust framework consists of 1) a domain-specific GAN model that can generate fake control/status messages, and 2) a probabilistic linear model with regression mechanism to realize risk and defense against attack surface on PGSC.

		\section{Zero Trust Framework Design}
		\begin{figure}[!t]
			\centerline{\includegraphics[scale=.5]{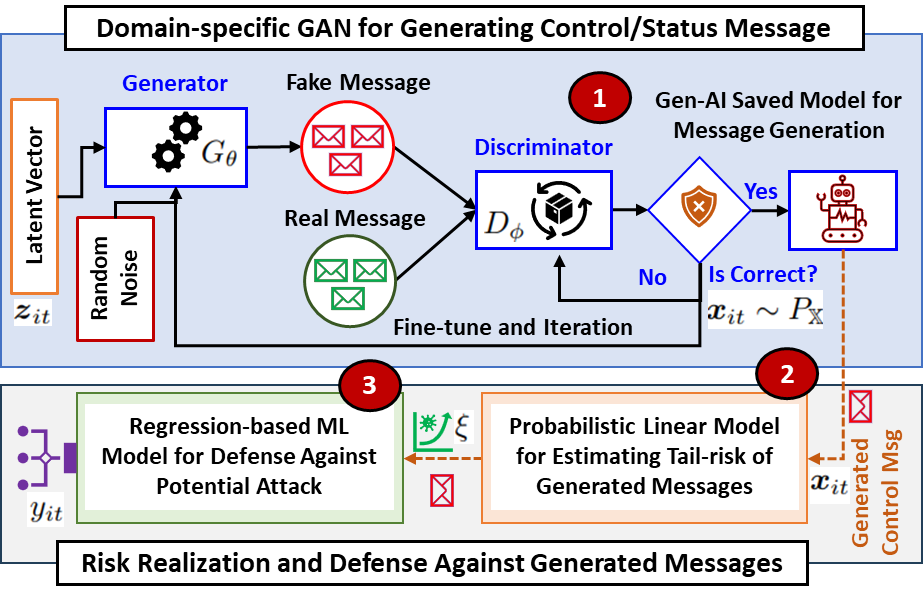}}
			\caption{Proposed zero trust framework for risk realization and defense against GenAI-driven attacks on the PGSC.}
			\label{sol}
		\end{figure}
		\begin{algorithm}[t!]
			\caption{GAN-based Training Algorithm for Control/Status Messages Generation in PGSC}
			\label{alg:gan}
			\begin{algorithmic}[1]
				\renewcommand{\algorithmicrequire}{\textbf{Input:}}
				\renewcommand{\algorithmicensure}{\textbf{Output:}}
				\REQUIRE  $\mathcal{I}$, $\mathbb{X}$
				\ENSURE  $\forall \boldsymbol{x}_{it} \in \mathcal{I}$
				\\ \textbf{Initialization}: $G_\theta$, $D_\phi$, $\theta$, $\phi$, $\boldsymbol{z}$ 
				\FOR {\textbf{Until max epoch:} $n \ge N$}
				\STATE \textbf{Mini batch:} $\mathbb{X}$, $\boldsymbol{z}$ 
				\STATE \textbf{Gradient decent $\theta$:}  $\bigtriangledown_{\theta} U (G_\theta, D_\phi)$ in \eqref{eq:decent_gen}  
				\STATE \textbf{Gradient ascent $\phi$:} $\bigtriangledown_{\phi} U (G_\theta, D_\phi)$ in \eqref{eq:ascent_dis}
				\STATE \textbf{Execute:} $\hat{D}_\phi(\boldsymbol{x}_{it}| G_{\theta}) \ge \frac{P_{\mathbb{X}} (\boldsymbol{x}_{it})}{P_{\mathbb{X}} (\boldsymbol{x}_{it}) + P_{G} (\boldsymbol{x}_{it})}$, in \eqref{Opt_1_1:const2}
				\ENDFOR
				\STATE \textbf{Trained model saved as $h5$ file}
				\RETURN $\theta$, $\phi$, $\boldsymbol{x}_{it}$
			\end{algorithmic} 
		\end{algorithm}
		We solve the formulated zero trust risk realization problem \eqref{Opt_1_1} by designing an analytical framework (as seen in Figure \ref{sol}) that can generate fake control/status messages, capable of quantifying extreme risk on generated messages, and protects the PGSC by autonomously detecting fake messages. In particular, we develop a domain-specific GAN mechanism to create the new attack vector by generating control/status messages $\boldsymbol{x}_{it}$ of DERs $\forall i \in \mathcal{I}$. We determine conditional-value-at-risk confidence level $\xi$ of the GenAI-driven attack vector by solving a probabilistic model while a regression-based machine learning (ML) model is devised to detect the fake $ y_{it}$ control message to protect PGSC.
		\subsection{A GAN for Producing New Attack Vector on PGSC}
		\begin{algorithm}[t!]
			\caption{Probabilistic and Regression-based Algorithm for Realizing Risk and Defense Against GenAI Attacks}
			\label{alg:sol_2}
			\begin{algorithmic}[1]
				\renewcommand{\algorithmicrequire}{\textbf{Input:}}
				\renewcommand{\algorithmicensure}{\textbf{Output:}}
				\REQUIRE  $\mathcal{I}$, $\eta$, $\boldsymbol{x}_{it}$, $\theta$, $\phi$, trained model (h5)
				\ENSURE  $\xi$,  $y_{it}$
				\\ \textbf{Initialization}: $\Theta_{i}$, $\beta_i$, $\Phi_i$, $\eta$, $\theta$, $\phi$ 
				\FOR { $t \ge T$}
				\FOR { $P(\Upsilon(\boldsymbol{x}_{it},\boldsymbol{z})) \ge \xi_{\eta} (\boldsymbol{x}_{it})$}  
				\STATE \textbf{Estimate: $\sigma$, $\mu$:} $g (\boldsymbol{x}_{it}) =  \frac{1}{\sigma \sqrt{2 \pi}} \exp \frac{(\boldsymbol{x}_{it} - \mu)^2}{2 \sigma^{2}}$
				\STATE \textbf{Estimate:} $\xi_{\eta} (\boldsymbol{x}_{it})$ = $\Gamma(1-\eta) * \sigma - \mu$ for \eqref{eq:var_eq}
				\STATE \textbf{Estimate:} $\Psi_{\eta}(\boldsymbol{x}_{it})$ = $\frac{1}{(1-\eta)} * \Omega(\xi_{\eta} (\boldsymbol{x}_{it})) * (\sigma - \mu)$ for \eqref{eq:cvar_eq}
				\STATE \textbf{Check:} Constraints \eqref{Opt_1_1:const1}, \eqref{Opt_1_1:const3}, \eqref{Opt_1_1:const4} and \textbf{Estimate:} $s_i(t)$ using \eqref{eq:Oscillator_model_with_freq_delay}
				\STATE \textbf{Estimate:} $\Lambda_{\eta}(\boldsymbol{x}, \xi)$ for \eqref{eq:var_cvar_eq_aux}
				\FOR { $i \ge |\mathcal{I}|$ $\&\& \; l$ }
				\STATE \textbf{Estimate:}  $y_{it}= \omega_0 + \omega_1 z_{1i} +  \dots + \omega_N z_{Ni}, \forall z_{Ni} \in \boldsymbol{z}_{it}$ for \eqref{Opt_1_1:const5} using bagging \cite{breiman1996bagging}
				\STATE \textbf{Check:} Constraint \eqref{Opt_1_1:const6}
				\ENDFOR
				\ENDFOR
				\ENDFOR
				\RETURN $\xi$,  $y_{it}$
			\end{algorithmic} 
		\end{algorithm}

		Algorithm \ref{alg:gan} illustrates the proposed GAN-based training mechanism for producing new attack vectors by generating DERs control/status messages on PGSC. We initialize a generator $G_\theta$, discriminator $D_\phi$, noise vector $\boldsymbol{z}$, learning parameters $\theta$ and $\phi$ at the beginning of Algorithm \ref{alg:gan}. Algorithm \ref{alg:gan} is designed for offline training, and thus, line $1$ determines the maximum number of epochs $N$ and line $2$ represents a high-level step to usage of mini batch during training. In line $3$ of Algorithm \ref{alg:gan}, we execute a gradient decent $\bigtriangledown_{\theta} U (G_\theta, D_\phi)$ mechanism to determine the learning parameters $\theta$ for the generator $G_\theta$ and evaluating a control message generation loss. The gradient decent of generator $G_\theta$ is given by:
		\vspace{-2mm}
		\begin{equation} \label{eq:decent_gen}
			\begin{split}
				\bigtriangledown_{\theta} U (G_\theta, D_\phi) = \frac{1}{|\mathcal{I}|} \sum_{i=1}^{|\mathcal{I}|} \bigtriangledown_{\theta} \log (1 -  D_\phi(G_\theta (\boldsymbol{z}_{it}))).
			\end{split}
		\end{equation}
		Then, Algorithm \ref{alg:gan} executes gradient ascent $\bigtriangledown_{\phi} U (G_\theta, D_\phi)$ to determine the learning parameters $\phi$ of discriminator $ D_\phi$ in line $4$. The gradient ascent $ D_\phi$ of the discriminator is given by:
		\begin{equation} \label{eq:ascent_dis}
			\begin{split} 
				\frac{1}{|\mathcal{I}|} \sum_{i=1}^{|\mathcal{I}|} \bigtriangledown_{\phi} \big[\log D_\phi(\boldsymbol{x}_{it}) + \log (1 -  D_\phi(G_\theta (\boldsymbol{z}_{it})) ) \big].  
			\end{split}
		\end{equation}
		In line $5$ of Algorithm \ref{alg:gan}, the discriminator $D_\phi$ assigns the probability of generated message and evaluating for fine tuning. Finally, a trained GAN model is saved to realizing generated AI-driven attacks on PGSC. 
		\subsection{Probabilistic and Regression-based Extreme Risk Realization and Defense Mechanism}
		We develop a probabilistic and regression-based mechanism for realizing extreme risk and defense against GenAI-driven control message attacks in PGSC.
		Algorithm \ref{alg:sol_2} presents the overall solution procedure to analyze the CVaR and defense mechanism for new attack vectors on the PGSC generated by Algorithm \ref{alg:gan}. Therefore, Algorithm \ref{alg:sol_2} receives a trained model as an input from Algorithm \ref{alg:gan} and generated control message $\boldsymbol{x}_{it}$. Line $2$ of Algorithm \ref{alg:sol_2} ensures the iterative process continues until $P(\Upsilon(\boldsymbol{x}_{it},\boldsymbol{z})) \ge \xi_{\eta} (\boldsymbol{x}_{it})$ while line $3$ estimates mean $\mu$ and standard deviation $\sigma$ for measuring the reconstruction capabilities of generated control message $\boldsymbol{x}_{it}$.  
		We derive a probability point function (PPF) $\Gamma(1-\eta)$ and estimate $\xi_{\eta} (\boldsymbol{x}_{it})$ the distribution of generated control message risk \eqref{eq:var_eq} as follows (line 4 in  Algorithm \ref{alg:sol_2}): 
		\vspace{-2mm}
		\begin{equation} \label{eq:var_eq_sol}
			\begin{split}
				\xi_{\eta} (\boldsymbol{x}_{it}) = \Gamma(1-\eta)  (\sigma - \mu),
			\end{split}
		\end{equation}
		where $\Gamma(1-\eta)$ is a probability point function and $\eta \in (0,1)$. Then, we construct a probability density function (PDF) $\Omega$ in line $5$ of Algorithm \ref{alg:sol_2} and capture the conditional expectation of CVaR for the AI generated controlled message $\boldsymbol{x}_{it}$. Thus, line $5$ of Algorithm \ref{alg:sol_2} execute the following function,   
		\begin{equation} \label{eq:cvar_eq_sol}
			\begin{split}
				\Psi_{\eta}(\boldsymbol{x}_{it}) = \frac{1}{(1-\eta)} * \Omega(\xi_{\eta} (\boldsymbol{x}_{it}))  \sigma - \mu,
			\end{split}
		\end{equation}
		where $\Omega(\xi_{\eta} (\boldsymbol{x}_{it}))$ is a PDF of generated controlled message $\boldsymbol{x}_{it}$. In Algorithm \ref{alg:sol_2}, line $6$ executes constraints \eqref{Opt_1_1:const1}, \eqref{Opt_1_1:const3}, and \eqref{Opt_1_1:const4} and estimates PGSC stability index $s_i(t)$ using \eqref{eq:Oscillator_model_with_freq_delay}. Line $7$ calculates the extreme risk (i.e., CVaR confidence level) cut-off point $\xi$ of the AI generated attack vector $\boldsymbol{x}_{it}$. Finally, lines $8$ to $11$ are responsible to distinguish between real $y_{it} = 0$ and generated $y_{it} = 1$ control messages $x_{it}$ to protect the PGSC from generated AI-driven attacks. The above solution provides a sub-optimal solution and performance relies on the parameter $\eta \in (0,1)$. 
		
		The complexity of the proposed zero trust risk realization and defense framework on PGSC completely depends on the complexity of Algorithm \ref{alg:sol_2} since Algorithm \ref{alg:sol_2} will be deployed in SCADA and being up and running. On the other hand, Algorithm \ref{alg:gan} is used for offline training to train an AI model for generating fake DER control/status messages while a trained model is being used by Algorithm \ref{alg:sol_2}. Therefore, the complexity of Algorithm \ref{alg:gan} can be ignored for the proposed zero trust framework on PGSC. Then, the complexity of Algorithm \ref{alg:sol_2} includes the complexity of two base problems: 1) a probabilistic linear model for extreme risk realization, and 2) a bagged-based random forest scheme for defense mechanism. Hence, the complexity of the probabilistic linear model-based risk realization becomes $\mathcal{O} (|\mathcal{I}|^2)$ \cite{Munir_CvaR1}, where $|\mathcal{I}|$ is the number of generated control messages of DERs $\forall i \in \mathcal{I}$. Now, we define $l$ as the number of bagged trees, where each message $\boldsymbol{x}_{it}$ consists of $|\boldsymbol{x}_{it}|$ features with the weight points $\omega$ during the regression learning for detecting AI generated control message. For a given number of bagged trees $l$, the overall complexity (i.e., time and space) of the defense mechanism belongs to $\mathcal{O} (l|\boldsymbol{x}_{it}|^2 |\boldsymbol{\omega}|^2  \log(|\boldsymbol{\omega}|))$, where $\mathcal{O} (l|\boldsymbol{x}_{it}| |\boldsymbol{\omega}|^2  \log(|\boldsymbol{\omega}|))$ is the time complexity. As a result, the total complexity of the proposed zero trust framework for PGSC leads to $\mathcal{O} (|\mathcal{I}|^2 + l |\boldsymbol{x}_{it}|^2 |\boldsymbol{\omega}|^2  \log(|\boldsymbol{\omega}|))$.              
		
		\section{Experimental Results and Analysis}
		\begin{table}[t!]
			\caption{Summary of Experimental Setup.}
			\begin{center}
				\begin{tabular}{|p{1.7cm}|p{5.8cm}|}
					\hline
					\textbf{Description} & \textbf{Values} \\
					\hline
					Generator & Sequential, $64$ units, ReLu (dense), Binary Cross-Entropy, Adam, LR: $0.02$, Latent Space: $35$, Epoch: $5000$  \\
					\hline
					Discriminator & Sequential, $64$ units, LeakyReLU (0.2) (dense), Sigmoid, Binary Cross-Entropy, Adam, LR: $0.02$, Latent Space: $35$, Epoch: $5000$   \\
					\hline
					RF bagging & estimators: $[50, 100, 200]$, max features: [auto, sqrt, log2], max depth: [2,4,5,6,7,8], criterion: [gini, entropy]    \\
					\hline
					CVaR & $\eta =\left\{0.9, 0.95, 0.99\right\}$ \\
					\hline
				\end{tabular}
				\label{exp_steup}
			\end{center}
		\end{table} 
		
		\begin{figure}[t!]
			\centering
			\begin{subfigure}[t]{0.45\textwidth}
				\centering
				\includegraphics[scale= 0.45]{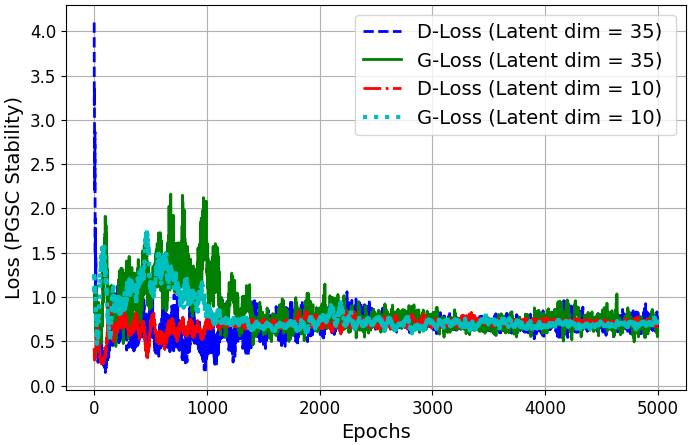}
				\caption{PGSC stability parameters.}
				\label{Loss_Stability}
			\end{subfigure}
			\\
			\begin{subfigure}[t]{0.45\textwidth}
				\centering
				\includegraphics[scale= 0.45]{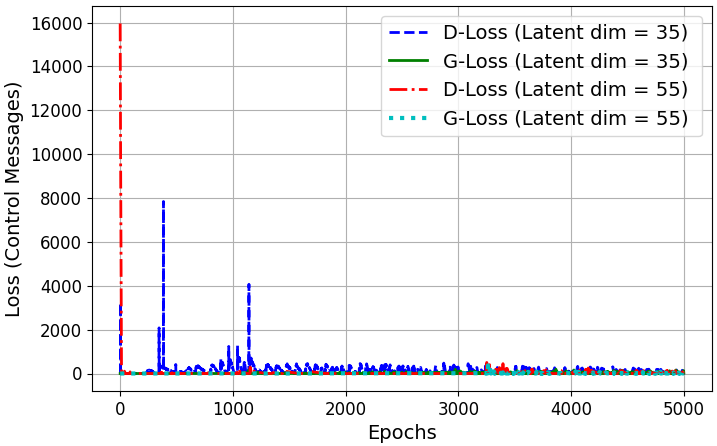}
				\caption{DERs control messages.}
				\label{Loss_Control_Msg}
			\end{subfigure}
			\caption{Generation and discrimination loss comparison of the proposed GAN-based model in PGSC.}
			\label{gan_con}
		\end{figure}
		
		\begin{figure}[t!]
			\centering
			\begin{subfigure}[t]{0.35\textwidth}
				\centering
				\includegraphics[scale= 0.35]{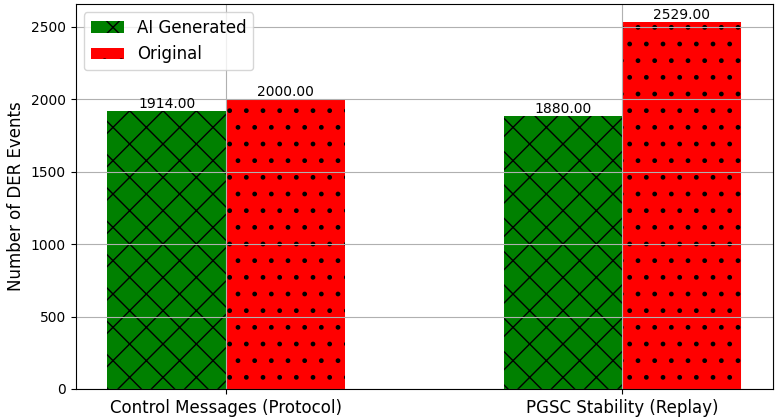}
				\caption{Protocol and replay attacks on PGSC by GenAI.}
				\label{gen_count}
			\end{subfigure}%
			\\
			\begin{subfigure}[t]{0.45\textwidth}
				\centering
				\includegraphics[scale= 0.45]{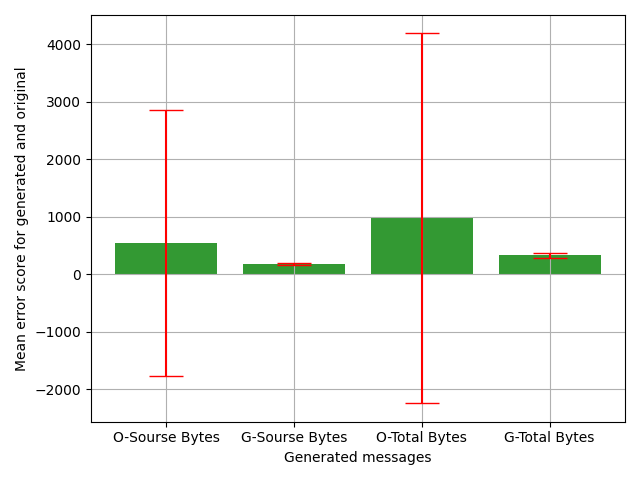}
				\caption{Error analysis of AI generated control message on PGSC.}
				\label{gen_msg}
			\end{subfigure}
			\caption{Capability of GenAI to create the attack vector on PGSC.}
			\label{error_acc}
			\vspace{-8mm}
		\end{figure}
		\begin{table}[t!]
			\caption{Performance analysis ($0$-$1$) for AI generated message detection among several regression-based models.}	
			\begin{center}
				\begin{tabular}{|c|c|c|c|c|}
							\hline
							\textbf{Methods} & \textbf{Precision} & \textbf{Recall} & \textbf{f1-score} & \textbf{Accuracy}   \\
							\hline
							RF (Bagging) & $1.0$ & $1$ & $1$  & $1.0$\\
							\hline
							KNN & $0.99$ & $1$ & $1$  & $0.99$\\
							\hline
							SVM  & $1.0$ & $1$ & $1$  & $1.0$\\
							\hline
							Logistic Regression  & $1.0$ & $1$ & $1$  & $1.0$\\
							\hline
						\end{tabular}
						\label{reliability}
					\end{center}
					\vspace{-6mm}
				\end{table} 
				\begin{figure}[t!]
					\centering
					\begin{subfigure}[t]{0.4\textwidth}
						\centering
						\includegraphics[scale= 0.45]{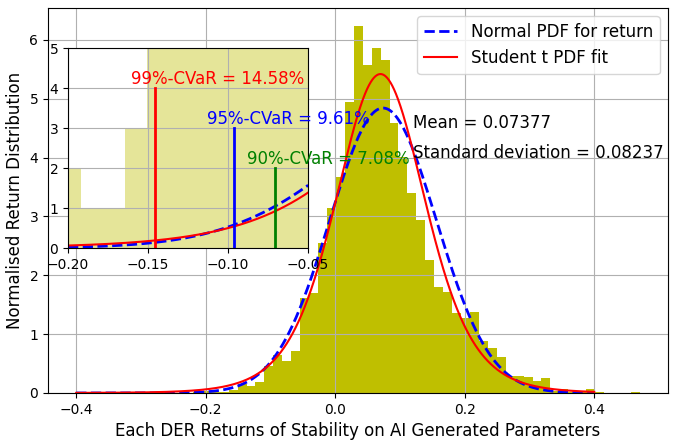}
						\caption{Replay attacks of PGSC.}
						\label{CVaR_Stability}
					\end{subfigure}%
					\\
					\begin{subfigure}[t]{0.4\textwidth}
						\centering
						\includegraphics[scale= 0.45]{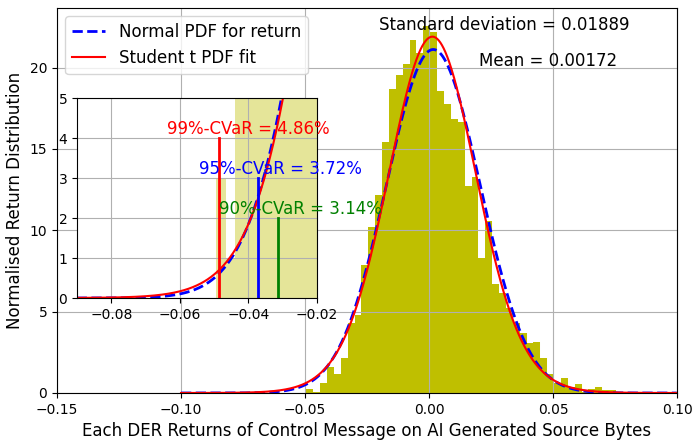}
						\caption{Protocol attacks of PGSC.}
						\label{CVaR_Control_Msg}
					\end{subfigure}
					\caption{Risk realization of AI-generated protocol and replay attacks on PGSC.}
					\label{cvar}
					\vspace{-2mm}
				\end{figure}
				
				The developed zero trust framework is one of the first work that attempts to realize and defense against GenAI-driven attacks on PGSC. Therefore, to the best of our knowledge, there are no prior works that can serve as a baseline. Therefore, we compare the proposed zero trust framework using two state-of-the-art datasets, 1) power grid stab stability \cite{UCI_Grid_Stability_Data}, and 2) SCADA control message \cite{WUSTL_data_2} to justify the efficacy. We summarize the important parameters of our experimental setup in Table \ref{exp_steup}.  
				
				In Figure \ref{gan_con}, we assess the convergence, generator loss, and discriminator loss of the proposed GAN-based training Algorithm \ref{alg:gan} for two datasets under different latent variables. We choose latent variable length as $35$ for both datasets (as seen in Figure \ref{Loss_Stability} for PGSC stability parameters and Figure \ref{Loss_Control_Msg} for control message generation) due to smooth convergence. Then, we analyze the capability of  creating new attack vector for both PGSC stability parameters and DER control message in Figure \ref{error_acc}, where we achieve around $95.7\%$ accuracy for protocol attack generation (in Figure \ref{gen_msg}) and about $74.3\%$ accuracy on replay attack generation (in Figure \ref{gen_count}).
				
				In Figure \ref{cvar}, we assess the extreme risk of the GenAI-driven protocol and replay attacks on the developed zero-trust framework. Figure \ref{CVaR_Stability} illustrates that the proposed framework can quantify the extreme risk $7.08\%$, $9.61\%$, and $14.58\%$ of GenAI-driven replay attacks for $90\%$, $95\%$, and $99\%$ confidence, respectively. Further, Figure \ref{CVaR_Control_Msg} demonstrates the extreme risk of GenAI-driven protocol attacks in PGSC, where the proposed framework can find $3.14\%$, $3.72\%$ and $4.86\%$ risk for $90\%$, $95\%$, and $99\%$ confidence, respectively. 
   
				In Table \ref{reliability}, we analyze the performance of the proposed bagging-based defense mechanism on zero trust framework over several regression-based methods. The results of Table \ref{reliability} clearly show that the proposed zero trust framework can effectively detect the GenAI-driven replay and protocol attacks on PGSC.

				\section{Conclusion}
				In this paper, we have introduced a novel zero-trust framework for the power grid to extreme risk realization and defense against generative AI-driven attacks such as protocol type and replay attacks on PGSC. In particular, we have designed the first approach to investigating GenAI-driven cyber attacks (i.e., protocol and replay) in PGSC, and created a novel zero-trust framework to realize and defend against GenAI attacks for PGSC.
				The proposed zero trust brings a state-of-the-art cybersecurity framework in the domain of critical power grid supply chains to protect the systems from AI-driven cyber attacks by continuously validating the trust of monitored DERs and their control messages.  
				Experimental results demonstrate the efficiency of the proposed zero trust framework, achieving an accuracy of $95.7\%$ in attack vector generation, a risk realization of $9.61\%$ for a $95\%$ stable PGSC, and a $99\%$ confidence level in defense against Generative AI-driven attacks. 
				In the future, we will further investigate the authentication of each DER to verify the data against being forged.

				\bibliographystyle{IEEEtran}
				\bibliography{referance.bib}

			\end{document}